\documentclass{article}

\usepackage{arxiv}

\usepackage[utf8]{inputenc} 
\usepackage[T1]{fontenc}    
\usepackage{hyperref}       
\usepackage{url}            
\usepackage{booktabs}       
\usepackage{amsfonts}       
\usepackage{nicefrac}       
\usepackage{microtype}      
\usepackage{lipsum}
\usepackage{graphicx}
\usepackage{subfigure}

\title{Improved Hybrid Layered Image Compression using Deep Learning and Traditional Codecs}

\author{
  Haisheng Fu \\
  School of Microelectronics\\
  Xi'an Jiaotong University\\
  \texttt{fhs4118005070@stu.xjtu.edu.cn} \\
   \And
  Feng Liang \\
  School of Microelectronics\\
  Xi'an Jiaotong University\\
  \texttt{fengliang@xjtu.edu.cn} \\
  \AND
  Bo Lei \\
  School of Microelectronics \\
  Xi'an Jiaotong University \\
  \texttt{lei\_bo@stu.xjtu.edu.cn} \\
   \And
  Nai Bian \\
  School of Microelectronics\\
  Xi'an Jiaotong University\\
  \texttt{biannai120@stu.xjtu.edu.cn} \\
   \And
  Qian zhang \\
  School of Microelectronics\\
  Xi'an Jiaotong University\\
  \texttt{zhangqian@stu.xjtu.edu.cn} \\
   \And
  Mohammad Akbari \\
  School of Engineering Science\\
  Simon Fraser University
  \texttt{akbari@sfu.ca}\\
  \And
  Jie Liang \\
  School of Engineering Science\\
  Simon Fraser University
  \texttt{ jiel@sfu.ca}\\
  \And
 Chengjie Tu\\
 Tencent Technologies\\
  \texttt{ chengjietu@tencent.com}\\
}

\begin{document}
\maketitle

\begin{abstract}
Recently deep learning-based methods have been applied in image compression and achieved many promising results. In this paper, we propose an improved hybrid layered image compression framework by combining deep learning and the traditional image codecs. At the encoder, we first use a convolutional neural network (CNN) to obtain a compact representation of the input image, which is losslessly encoded by the FLIF codec as the base layer of the bit stream. A coarse reconstruction of the input is obtained by another CNN from the reconstructed compact representation. The residual between the input and the coarse reconstruction is then obtained and encoded by the H.265/HEVC-based BPG codec as the enhancement layer of the bit stream. Experimental results using the Kodak and Tecnick datasets show that the proposed scheme outperforms the state-of-the-art deep learning-based layered coding scheme and traditional codecs including BPG in both PSNR and MS-SSIM metrics across a wide range of bit rates, when the images are coded in the RGB444 domain.
\end{abstract}


\section{Introduction}
Recently with the rapid development of deep learning theory, especially after the successful applications of convolutional neural networks (CNN) in computer vision, deep learning has been applied to many areas, including image compression. Some deep learning-based methods have outperformed traditional image codecs such as JPEG, JPEG2000, and the H.265/HEVC-based BPG image codec \cite{Bpg_125, NIPS2017_6714, DBLP:journals/corr/abs-1804-02958, Li_2018_CVPR, Johnston_2018_CVPR, 8456298, 8100060, Agustsson_2018_CVPR}, demonstrating its great  potentials. In \cite{DBLP:journals/corr/abs-1511-06085}, Toderici et al. proposed a variable-rate image compression framework for thumbnail images by using the recurrent neural networks (RNNs). The scheme was generalized in \cite{8100060} to compress images of any size and achieve better results than JPEG. The convolutional autoencoder (CAE) framework with a coding network and a decoding network was used for image compression in \cite{ICML2017_12}. In \cite{ICLR2017_67}, Belle et al. introduced a coding scheme with a generalized divisive normalization (GDN)-based nonlinear analysis transform, a uniform quantizer and a nonlinear synthesis transform. 

The presence of quantization, which is non-differentiable, in the pipeline of image coding has posed some challenges to deep learning-based schemes, which use gradient-based backpropagation to train neural networks. To solve the problem, some approximations of the quantization have been proposed. For example, in \cite{ICLR2017_28}, the quantization of the autoencoder is replaced by a smooth approximation in the backpropagation because the derivative of the rounding function is zero almost everywhere, but the forward pass of the backpropagation still use the quantization to prevent the decoder from learning to invert the smooth approximation. A soft (continuous) relaxation of quantization and entropy was developed in \cite{NIPS2017_6714}. The designed network is not used to give a specific quantized output, but is used to output a quantization level along with the weight.

The generative adversarial network (GAN) \cite{NIPS2014_5423} is a powerful and versatile scheme, and has also been adopted in image compression. In \cite{8456298}, a decoder was trained to generate realistic images using a discriminator. The similarity is measured by perceptual loss based on the feature map of a pretrained AlexNet. In \cite{DBLP:journals/corr/abs-1801-09468, DBLP:journals/corr/abs-1804-02958},  the feature maps were extracted from learning-based image compression to complete image segmentation or image classification. Based on the work in \cite{8456298, DBLP:journals/corr/abs-1801-09468, DBLP:journals/corr/abs-1804-02958}, the work \cite{DBLP:journals/corr/abs-1804-02958} used segmentation map-based image synthesis to train the discriminator of GAN, which used synthesized images for non-important regions and obtained very good performance at low bite rates ($<$0.1 bits/pixel). 

\begin{figure}
\centering
\subfigure[Original]{
\begin{minipage}[t]{0.25\linewidth}
\centering
\includegraphics[scale=0.8]{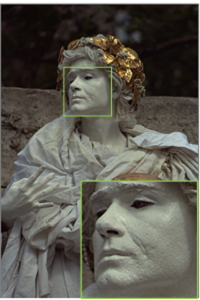}
\end{minipage}
}%
\subfigure[BPG (0.294 / 35.03 / 0.981)]{
\begin{minipage}[t]{0.25\linewidth}
\centering
\includegraphics[scale=0.8]{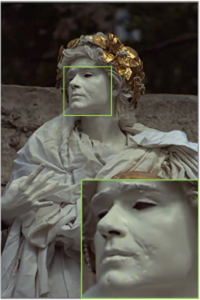}
\end{minipage}
}%
\subfigure[DSSLIC (0.283 / 36.02 / 0.983)]{
\begin{minipage}[t]{0.25\linewidth}
\centering
\includegraphics[scale=0.8]{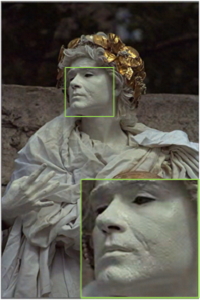}

\end{minipage}
}%
\subfigure[Ours (0.280 / 36.564 / 0.984)]{
\begin{minipage}[t]{0.25\linewidth}
\centering
\includegraphics[scale=0.8]{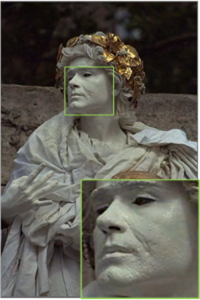}
\end{minipage}
}%
\centering
\caption{An example of the decoded images of the BPG, DSSLIC, and the proposed method (bits/pixel/channel, PSNR(dB), MS-SSIM).}
\label{FIG:1}
\end{figure}

In \cite{8683541}, a deep semantic segmentation-based layered image compression (DSSLIC) scheme is proposed, which is a hybrid coding approach that uses both deep learning and the traditional codecs such as the BPG and FLIF \cite{7532320}. It has three layers. First, it obtains the semantic map of the original image and encodes it by the FLIF codec as the base layer, which is used to synthesize an initial reconstruction. To improve the image synthesis quality, a low-dimensional representation of the input is generated and coded by FLIF as the second layer, which is served as a side information to the image synthesis. Moreover, the residual between the input and the synthesized image is encoded by the BPG codec as the third layer. The DSSLIC outperforms the BPG codec (in RGB444 format) by large margin in both PSNR and multi-scale structural similarity (MS-SSIM) metrics across a large range of bit rates.

Although the performance of DSSLIC is quite impressive, we observe that in many cases the semantic information in DSSLIC is not very useful. For example, when the semantic segmentation part fails to assign any label to a region, the image synthesis part is actually not used at all, and the remaining parts of the DSSLIC still works very well. This motivates us to develop a simplified hybrid coding scheme without using semantic segmentation-based image synthesis.

In this paper, we propose an improved hybrid image compression framework. Our scheme only has two layers. Its base layer is essentially an autoencoder, and the result is coded by the FLIF codec. The enhancement layer is the residual between the input and the reconstruction of the autoencoder, which is coded by the BPG codec. A number of contributions are made in the design of the autoencoder. We change the structure of the original residual block in the ResNet in \cite{He_2016_CVPR} and add the dropout module between two convolution layers to improve the generalization capability of the network. We also replace the default rectified linear units (ReLU) by the parametric rectified linear units (PReLU). Despite the simplified scheme, our method can outperform the DSSLIC and other traditional codecs in both PSNR and MS-SSIM metrics across large range of bit rates, especially in PSNR. An example is shown in Fig. \ref{FIG:1} by comparing with the DSSLIC and BPG. More results are reported in Sec. 4.

The organization of the paper is as follows. The framework and the details of the improved hybrid image compression method are introduced in Sec. \ref{Framework}. In Sec. 3, we evaluate the performance of our method by comparing with DSSLIC, BPG, JPEG, JPEG2000 and WebP codecs using the popular Kodak and Tecnick datasets in the RGB444 domain. Conclusions and discussions are given in Sec. 4.

\begin{figure}
	\centering
		\includegraphics[scale=.75]{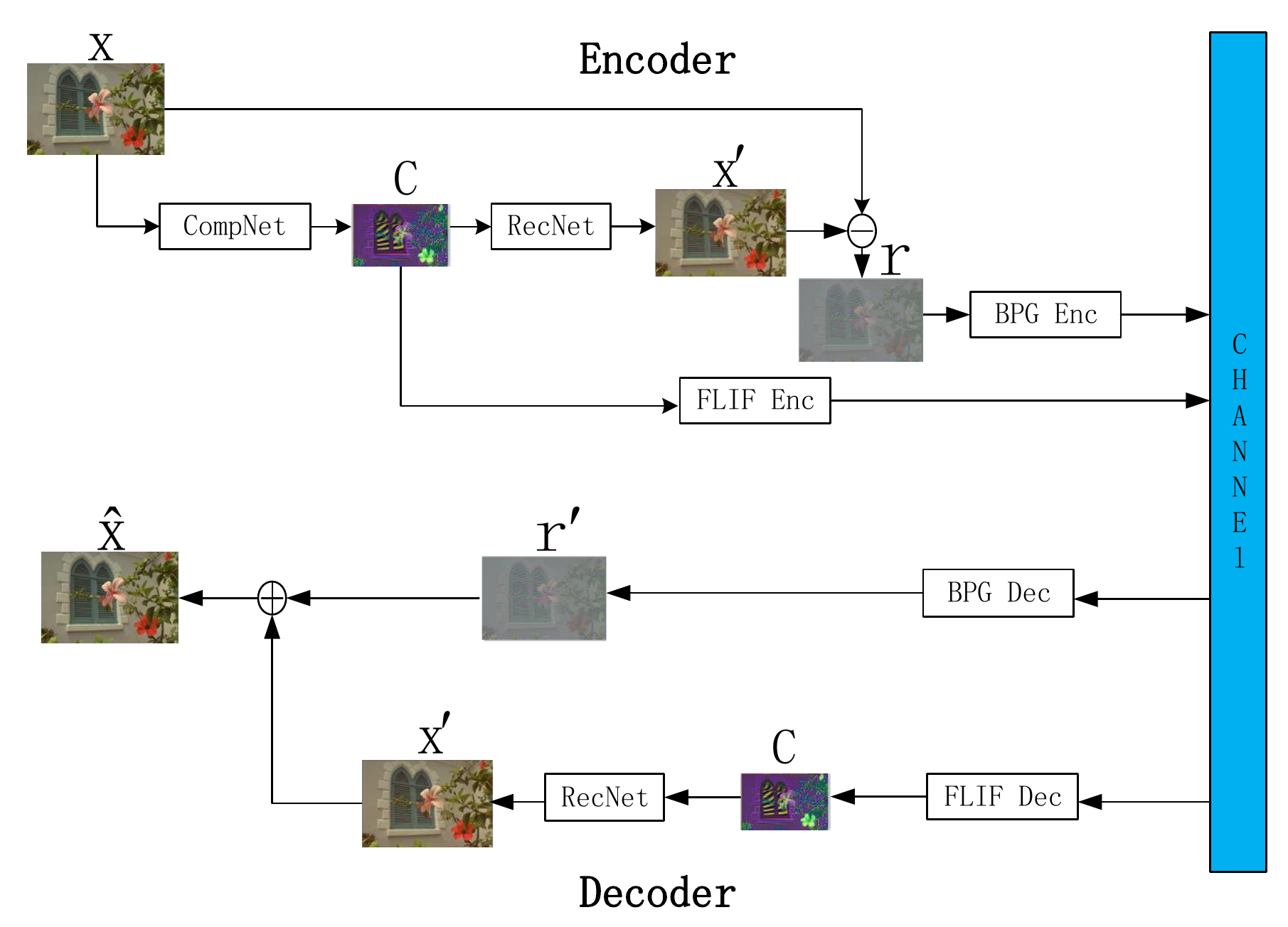}
	\caption{The Framework of the proposed improved hybrid image compression network. }
	\label{FIG:2}
\end{figure}

\begin{figure}
	\centering
		\includegraphics[scale=1.0]{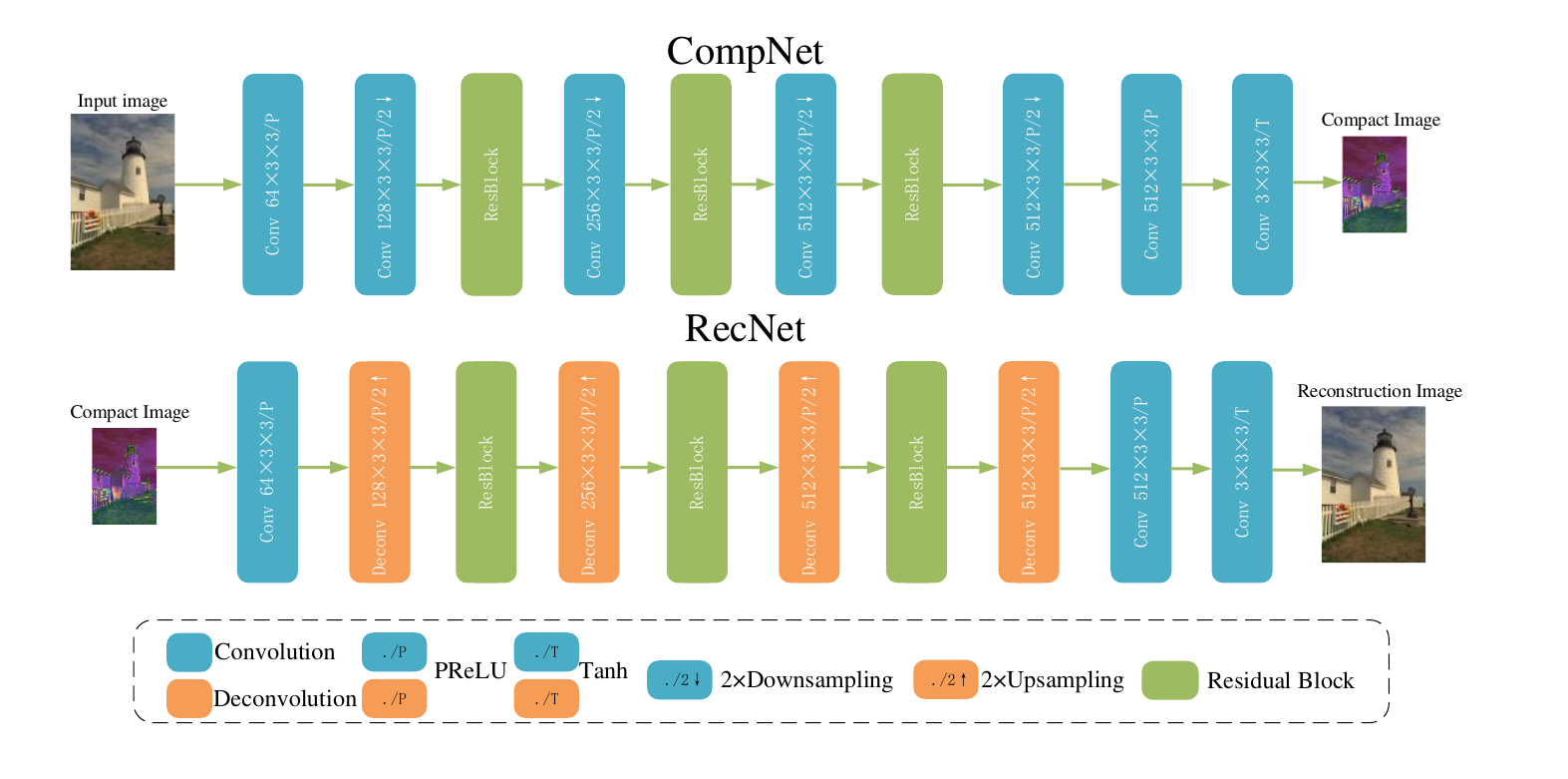}
	\caption{The structures of the CompNet and RecNet.}
	\label{FIG:3}
\end{figure}

\section{The Proposed Hybrid Image Compression Scheme}
\label{Framework}

\subsection{Framework}

The proposed framework is described in Fig. \ref{FIG:2}. It includes two layers, which is simpler than the DSSLIC. We first use a deep learning network named CompNet to learn the compact image $c$ of the original image $x$, which is losslessly encoded by the FLIF codec. Next, the compact image is used by another deep learning network named RecNet to produce a coarse reconstruction $x'$ of the input image. CompNet and RecNet thus form a convolutional autoencoder (CAE). After that, the residual $r$ between the input image and the coarse reconstruction image is obtained and encoded by the lossy BPG codec. In the decoder side, the compact image $c$ is decoded, which is used by the RecNet to obtain the coarse reconstruction $x'$. The residual image is decoded by the BPG decoder, denoted by $r'$. Finally, the coarse reconstruction $x'$ and the decoded residual image $r'$ are added to get the final reconstruction $\hat{x}$. 

\subsection{Network Architecture}

The network structures of the CompNet and RecNet are described  in Fig. \ref{FIG:3}.  The two networks constitute a complete CAE. Both of them have 13 network layers. The CompNet contains four downsampling layers and the RecNet contains four upsampling layers. There is a residual network which contains two convolution layers between the two upsampling or downsampling layers. The initial image size is $W\times H\times 3$. All the kernel sizes of the convolutional layers and deconvolutional layers are $3\times 3$.  The width and the height of the compact image $c$ are both $1/16$ of those of the input image.

In order to improve the performance of image compression, we add a residual block between every two upsampling layers and downsampling layers, motivated by the skip connection in the ResNet \cite{He_2016_CVPR}, which can deepen our network and also achieve faster convergence. The work \cite{ICLR2017_67, ICLR2018_26} introduce the generalized divisive normalization (GDN) and inverse GDN (IGDN) modules, based on a local divisive normalization transformation. Compared with the residual blocks, the networks that rely on GDN/IGDN modules may have vanishing gradient and exploding gradient problems during training as the increase of the number of network layers. Since our goals are to obtain compact images of smaller size and reconstructed images of higher quality, which call for deeper networks. Therefore we choose to use the residual block as the basic module of our networks.

\begin{figure}
	\begin{center}
	\includegraphics[width=\columnwidth]{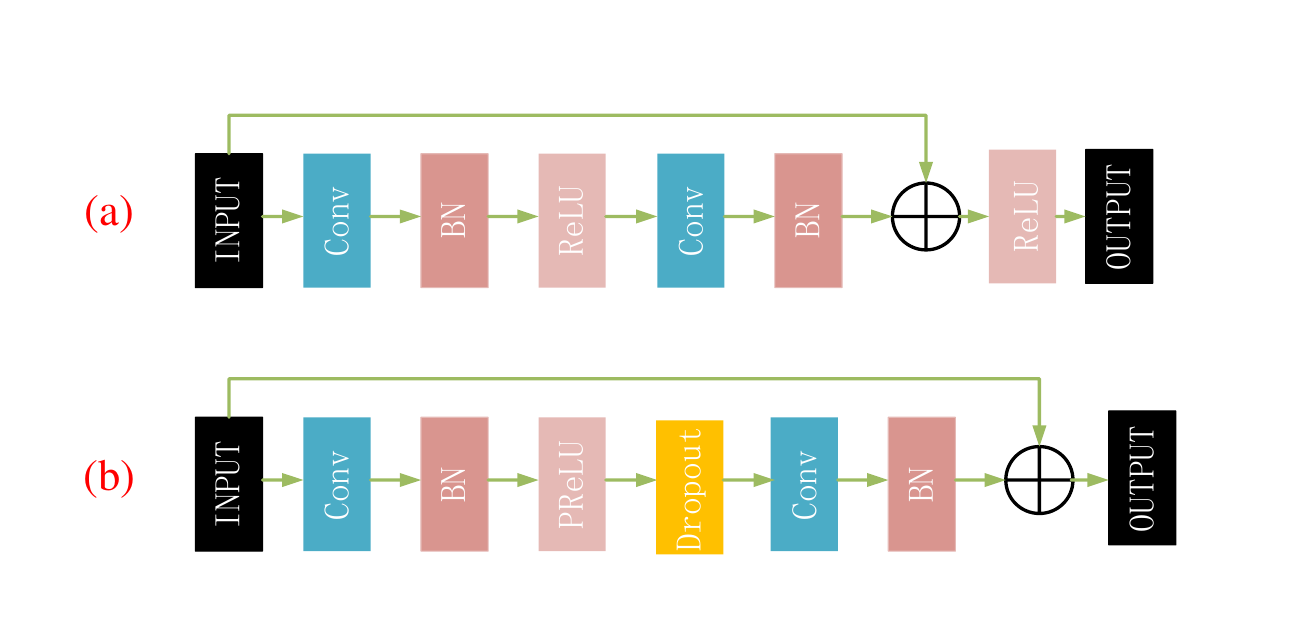}
		\caption{(a) The original residual block in \cite{He_2016_CVPR}. (b) The proposed residual block.}
	\label{FIG:4}
	\end{center}
\end{figure}

In the residual block, we use two $3\times 3$ convolution layers with $k$ filters. Inspired by \cite{DBLP:journals/corr/abs-1806-01496}, we also make some changes to the structure of the residual block. We remove the nonlinear mapping after the skip connection for each residual unit and add the dropout module between the two $3\times 3$ convolution layers. Experimental results show that this can speed up the convergence during training, and can prevent over-fitting as well. The detailed description is shown in Fig. \ref{FIG:4}.

In addition, inspired by \cite{DBLP:journals/corr/abs-1806-01496, NIPS2018_8275}, we replace the rectified linear units (ReLU) by the parametric rectified linear units (PReLU), which can achieve a faster convergence speed than ReLU \cite{DBLP:journals/corr/abs-1806-01496}.  In the experimental section, we will investigate the effects of these two activation functions on image compression performance. 

Moreover, in the last layers of CompNet and RecNet, we use the Tanh activation function to make the output range distributed in [-1,1].

\subsection{Scaling in the System}

In our system, the data at different stages have different ranges. The pixel values of the input image are 8-bit unsigned integers in the range of $[0, 255]$. First, the inputs to the CompNet should be normalized to the range of $[-1, 1]$ as follows. 
\begin{equation}
    x_f = \frac{x}{255} \times 2 - 1.
    \label{scale_i2f}
\end{equation}

The output of CompNet is also in $[-1, 1]$. In order to encode the output of the CompNet by the FLIF codec, we scale it to 8-bit integers in $[0, 255]$ as follows, so that it can be accepted by FLIF as a valid image.
\begin{equation}
	x_Q=round[\frac{x_f+1}{2} \times 255].
	\label{scale_f2i}
\end{equation}

The scaled result $x_Q$ is then normalized to $[-1, 1]$ by Eq. \ref{scale_i2f}, and is sent as the input to the ResNet. The output of ResNet is also in the range of $[-1, 1]$, and is converted to 8-bit integers using Eq. \ref{scale_f2i}. This is the coarse reconstruction of the image.

The residual image $r$ is the difference between the original image and the coarse reconstruction. Therefore its range is $[-255, 255]$. Since we use the traditional BPG codec to encode the residual, another scaling function is needed to map the residual image pixels from $[-255, 255]$ to $[0, 255]$, so that the residual image can be accepted by the BPG codec as a valid input image.

In \cite{8683541} and its early version, two mapping methods are used. The first one is the following simple shift method:
\begin{equation}
  x' \leftarrow round[\frac{x'+255}{2}].
  \label{eq_shift}
\end{equation}

A better scaling method is to find the maximum and minimum values of the residual images, $x'_{max}$ and $x'_{min}$, and then adaptively map the range between them to [0, 255]. That is:
\begin{equation}
  x' \leftarrow round[\frac{(x'-x'_{min})}{(x'_{max}-x'_{min})} \times 255].
  \label{eq_minmax}
\end{equation}

In this method, the minimum and maximum values need to be sent to the decoder to do the inverse scaling, but the overhead is negligible. The shift method in Eq. \ref{eq_shift} is a special case of this adaptive method when the maximum and minimum values are 255 and -255 respectively. 

\begin{figure}
	\begin{center}
    	\includegraphics[width=\columnwidth]{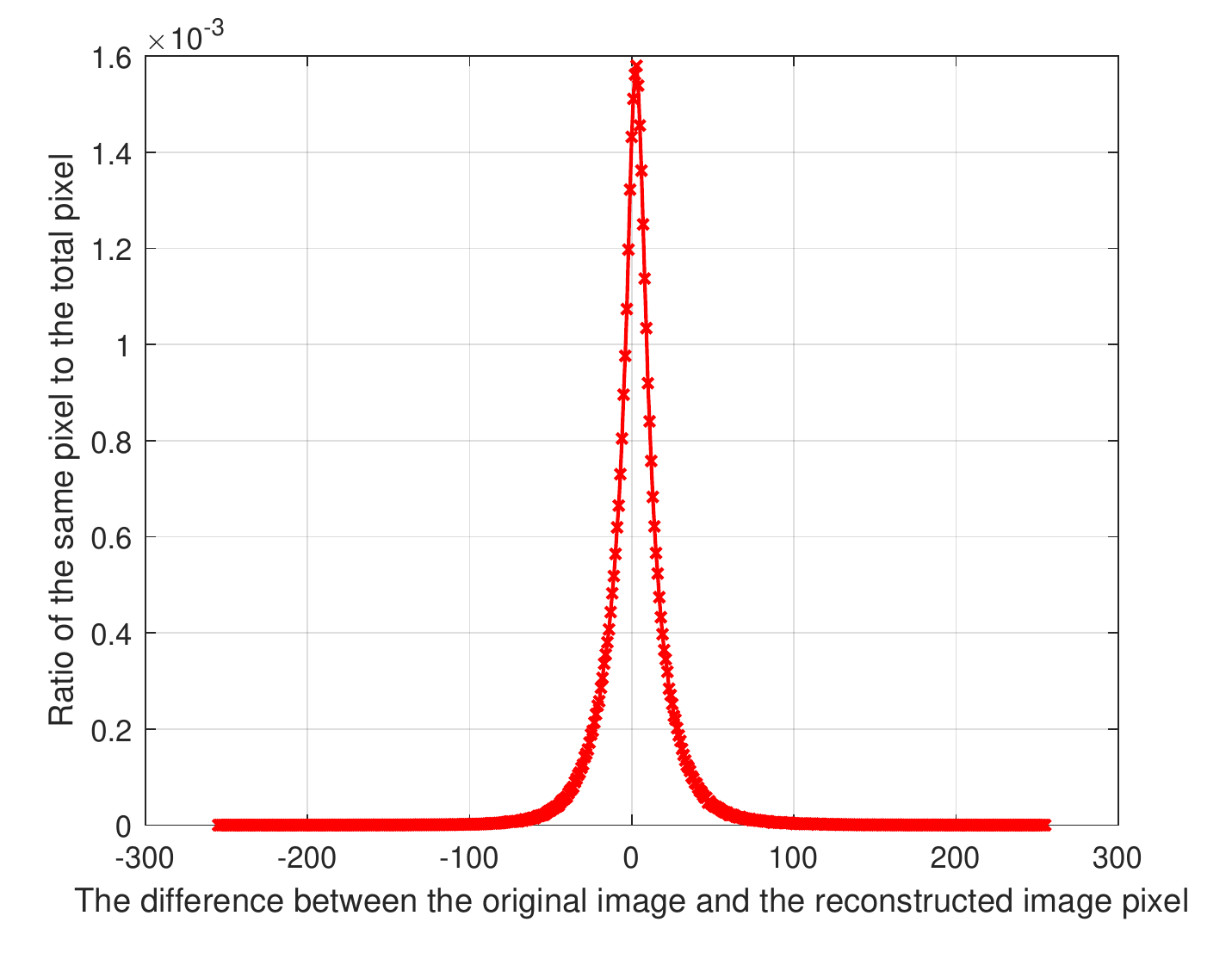}
		\caption{Distribution of the residual image pixels between the original image and the coarse reconstruction.}
    	\label{res_dist}
	\end{center}
\end{figure}

Our experimental results show that in most cases, the residual image pixel values generally obey a normal distribution with mean around zero and a very small variance. That is, many residuals are very close to 0, and most residuals are within a small range, as shown by the example in Fig. \ref{res_dist}, hence the adaptive method can preserve more details of the residual, and yields better performance than the simple shift method. 

However, sometimes the maximum and minimum are quite close to $255$ and $-255$, making the adaptive scaling method very similar to the shift method, although most residuals are very small, and there are only a few pixels with extremely large absolute values. 

To address this problem, we propose a clipping method. It can be seen from Fig. \ref{res_dist} that most of the values fall in $[-120, 120]$. Therefore we set the maximum value to $120$ and the minimum value to $-120$ in Eq. \ref{eq_minmax}. Extreme residual pixels beyond $[-120, 120]$ are clipped to $-120$ or $120$. This can reduce the range and preserve more details after the scaling. We call this the clipping method. We will show later that the clipping method is slightly better than the adaptive method, and the latter is better than the simple scaling method.

\subsection{Network Training}

We use the CLIC compression dataset released by the Computer Vision Lab of ETH Zurich to train our network. In order to train our network better, we use several data augmentation methods such as rotation and scaling to randomly extract 81,650 patches from 1633 high-quality images. These images are saved as lossless PNGs to avoid compression artifacts. We only use the mean squared error (MSE) as the loss function to train our network.
\begin{equation}
	L=\frac{1}{N} \left | Y_{n}-X_{n} \right |^2,
\end{equation}
where $X_{n}$ represents the initial input image, and $Y_{n}$ represents the coarse reconstruction, which is the output of the RecNet. $N$ is the batch size. In order to improve the SSIM performance of the reconstructed images, many papers have added some other loss functions such as VGG Loss, Perception Loss, and Adversarial Loss in the training process. Although these loss functions can improve the MS-SSIM metric of the reconstructed images, they will degrade the PSNR of the image. We will show later that although the SSIM is not considered in our cost function, our method still outperforms other methods in SSIM.

The CompNet and RecNet are jointly trained for 40 epochs with mini-batch stochastic gradient descent method (SGD). The batch size is set to 20. The Adam solver with fixed learning rate of 0.0001 is used.

\section{Experiments}

In this section, we evaluate the performance of our method on the Kodak PhotoCD dataset\cite{Kodak_dataset} and Tecnick dataset \cite{Tecnick_dataset}, and compare with some other methods, including DSSLIC, BPG, JPEG2000, WebP, and JPEG. All images are coded by the RGB444 format in all codecs. We use both the PSNR and MS-SSIM metrics to compare the performance of different codecs. 

In our method, we evaluate three scaling methods of the residual images, namely, the simple shift method in Eq. \ref{eq_shift}, the MinMax method in Eq. \ref{eq_minmax}, and the Clipping method.

\subsection{Image Coding Results}

\begin{figure}
\centering
\subfigure{
\begin{minipage}[t]{0.5\linewidth}
\centering
\includegraphics[width=\columnwidth]{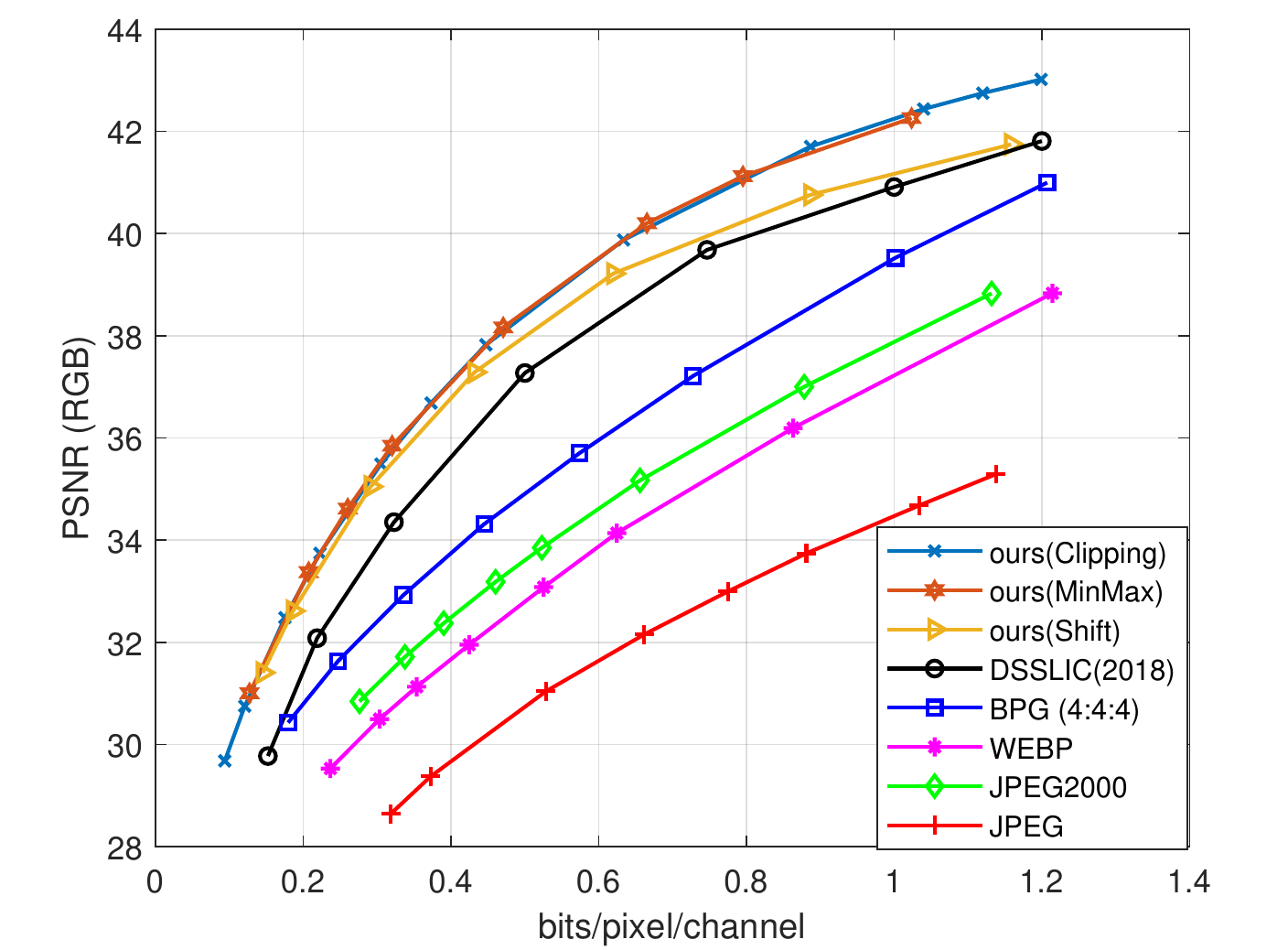}
\end{minipage}
}%
\subfigure{
\begin{minipage}[t]{0.5\linewidth}
\centering
\includegraphics[width=\columnwidth]{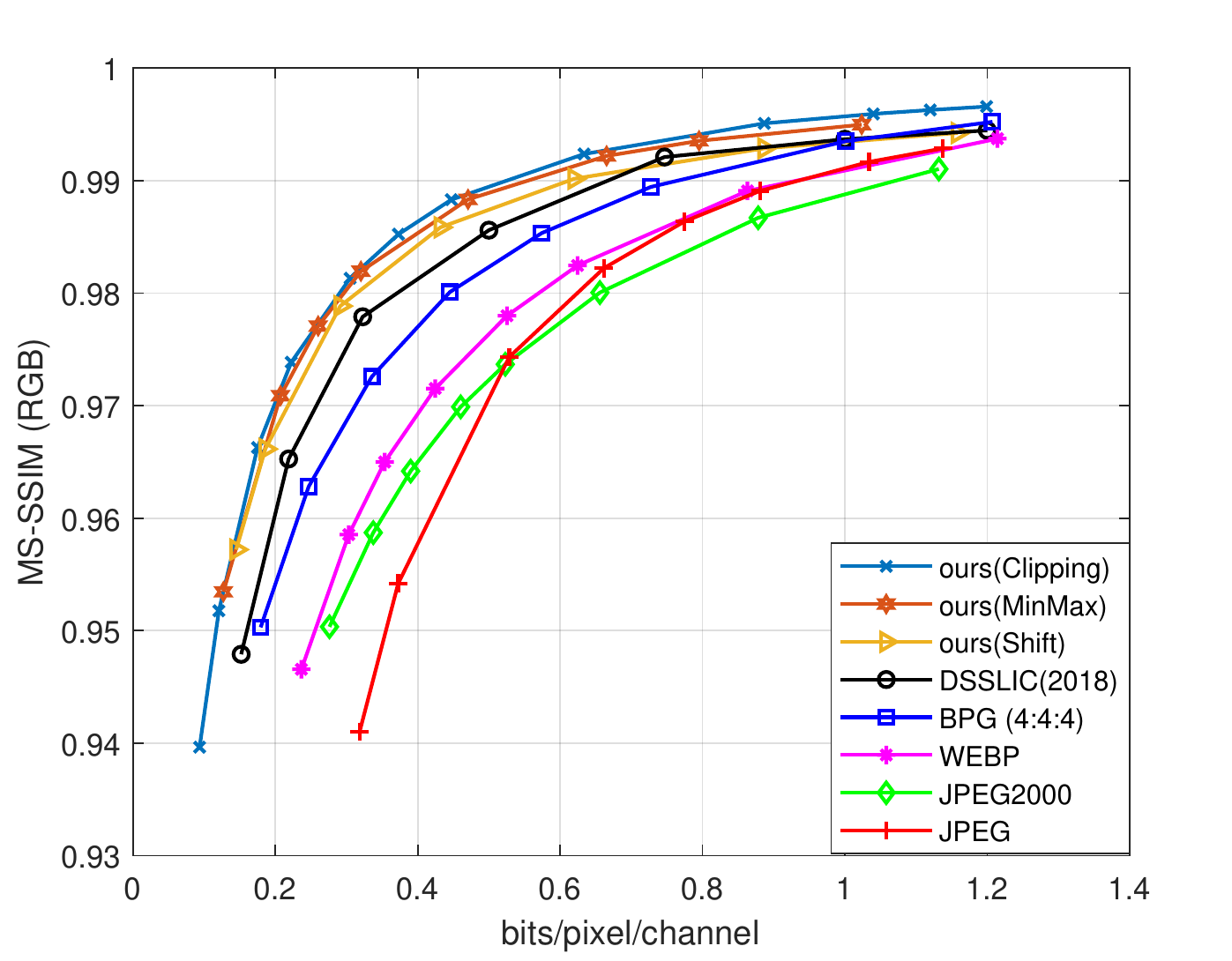}
\end{minipage}
}%
\centering
\caption{Average comparison results on all 24 Kodak images in terms of PSNR and MS-SSIM.}
\label{test_kodakdataset}
\end{figure}

The average PSNR performance and MS-SSIM performance over the 24 Kodak images are illustrated in Fig. \ref{test_kodakdataset}. All three versions of our method achieve better results than other methods in both PSNR and MS-SSIM. The Clipping method and the MinMax method have almost the same performance, and consistently outperform the DSSLIC, with and average gain of over 1 dB. The Shift method has similar performance as the other scaling methods at low rates, but the performance gets worse at high rates. However, it still outperforms the DSSLIC. The DSSLIC outperforms BPG and other codecs by a large margin.

\begin{figure}
\centering
\subfigure{
\begin{minipage}[t]{0.5\linewidth}
\centering
\includegraphics[width=\columnwidth]{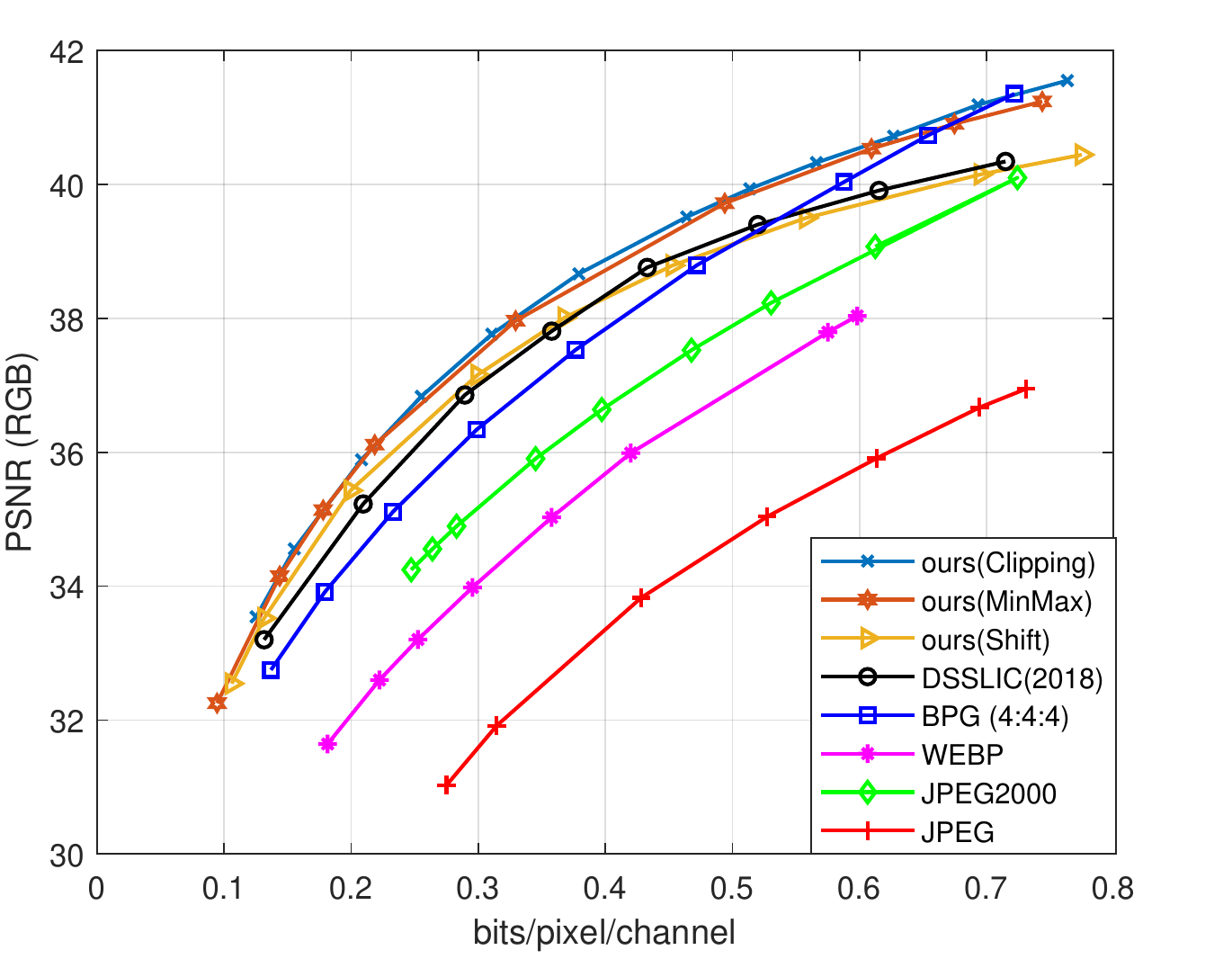}
\end{minipage}
}%
\subfigure{
\begin{minipage}[t]{0.5\linewidth}
\centering
\includegraphics[width=\columnwidth]{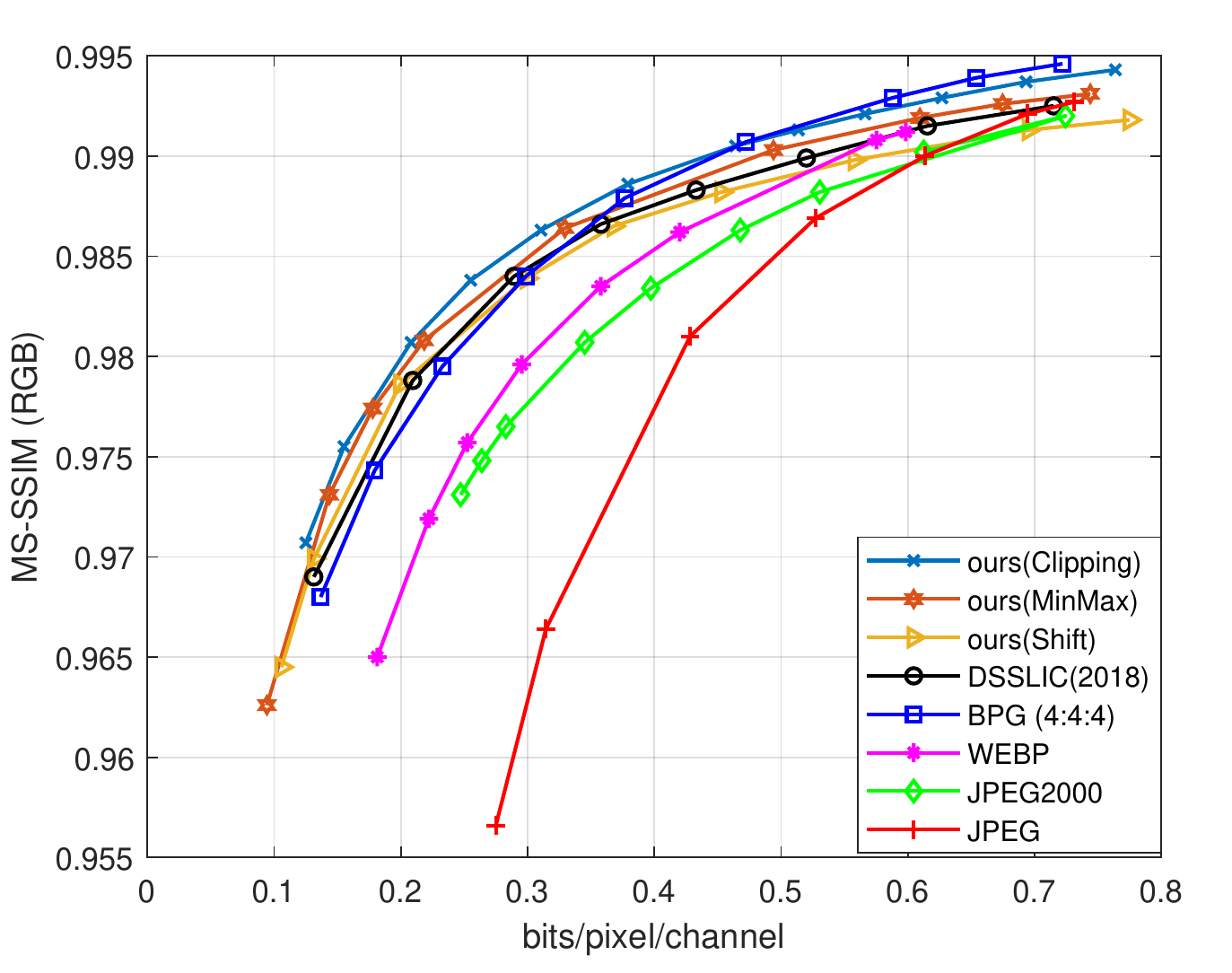}
\end{minipage}
}%
\centering
\caption{Average comparison results on all 100 Tecnick images in terms of PSNR and MS-SSIM.}
\label{test_Tecnickdataset}
\end{figure}

Fig. \ref{test_Tecnickdataset} compares the performance of our method with others on the Tecnick dataset. In PSNR, the Clipping method and the MinMax method still outperform DSSLIC and BPG. The Shift method is similar to DSSLIC, and is worse than BPG at high rates. In the MS-SSIM metric, the clipping and MinMax methods are always better than DSSLIC. They are also better than BPG at low rates. The shift method is slightly worse than DSSLIC.

\begin{figure}
\centering
\subfigure[Original]{
\begin{minipage}[t]{0.33\linewidth}
\centering
\includegraphics[scale=0.29]{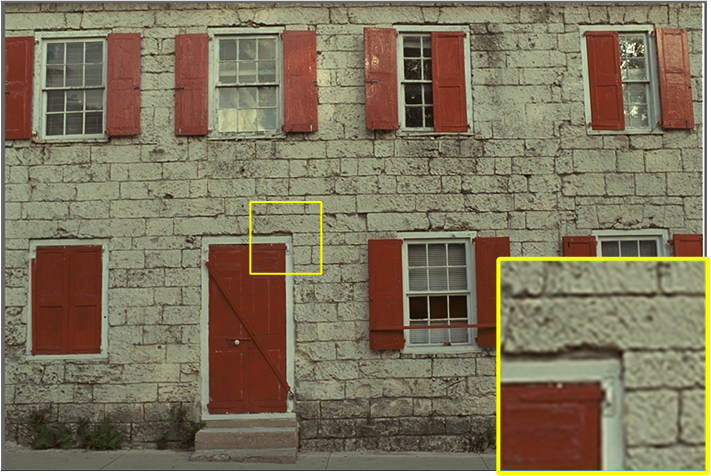}
\end{minipage}
}%
\subfigure[JPEG(0.19/22.5/0.817)]{
\begin{minipage}[t]{0.33\linewidth}
\centering
\includegraphics[scale=0.29]{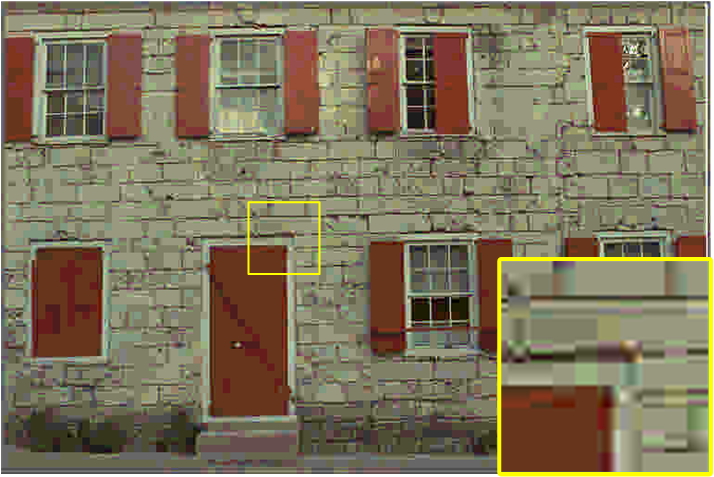}
\end{minipage}
}%
\subfigure[JPEG2000(0.218/25.1/0.903)]{
\begin{minipage}[t]{0.33\linewidth}
\centering
\includegraphics[scale=0.29]{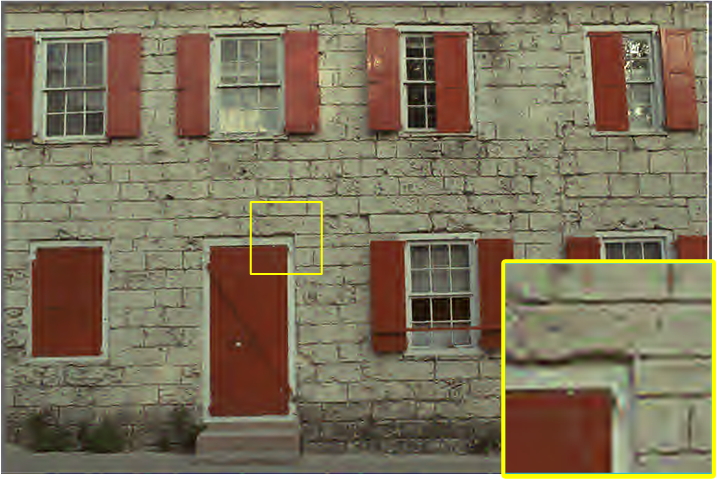}
\end{minipage}
}%

\subfigure[BPG(0.190/26.3/0.930)]{
\begin{minipage}[t]{0.33\linewidth}
\centering
\includegraphics[scale=0.29]{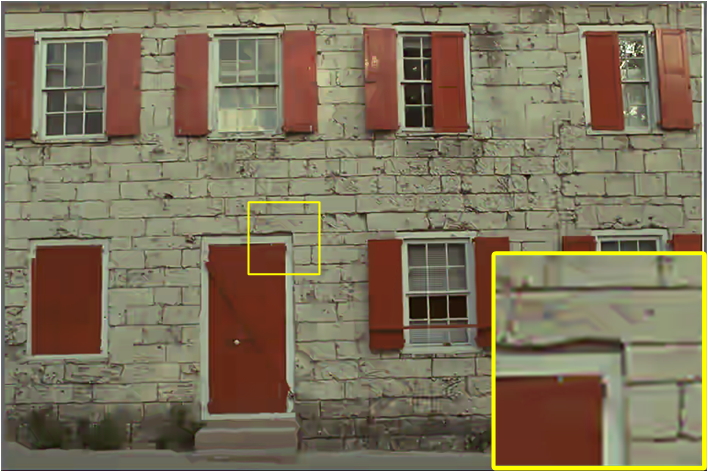}
\end{minipage}
}%
\subfigure[DSSLIC(0.184/27.0/0.940)]{
\begin{minipage}[t]{0.33\linewidth}
\centering
\includegraphics[scale=0.29]{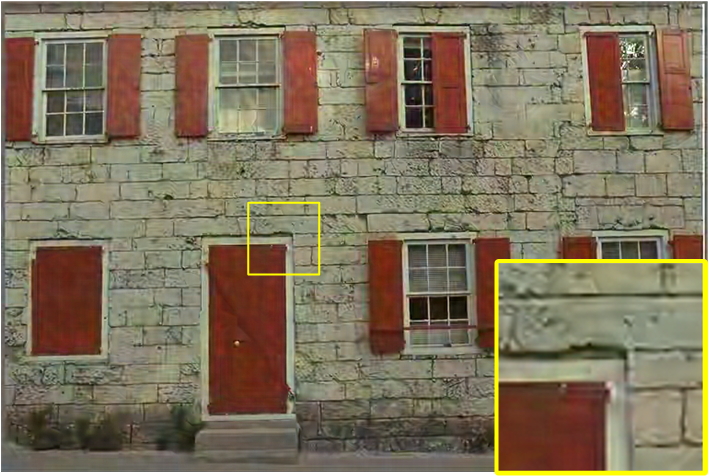}
\end{minipage}
}%
\subfigure[Ours(0.182/27.8/0.949)]{
\begin{minipage}[t]{0.33\linewidth}
\centering
\includegraphics[scale=0.29]{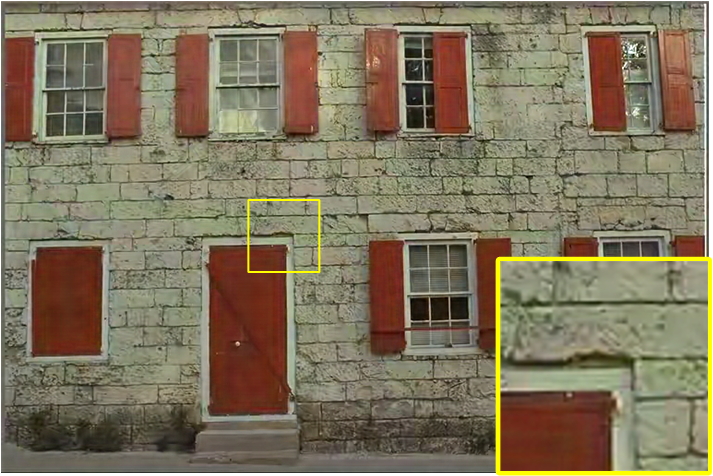}
\end{minipage}
}%
\centering
\caption{Example 1 in the Kodak dataset (bits/pixel/channel, PSNR, MS-SSIM).}
\label{Example_1}
\end{figure}

\begin{figure}
\centering
\subfigure[Original]{
\begin{minipage}[t]{0.33\linewidth}
\centering
\includegraphics[scale=0.25]{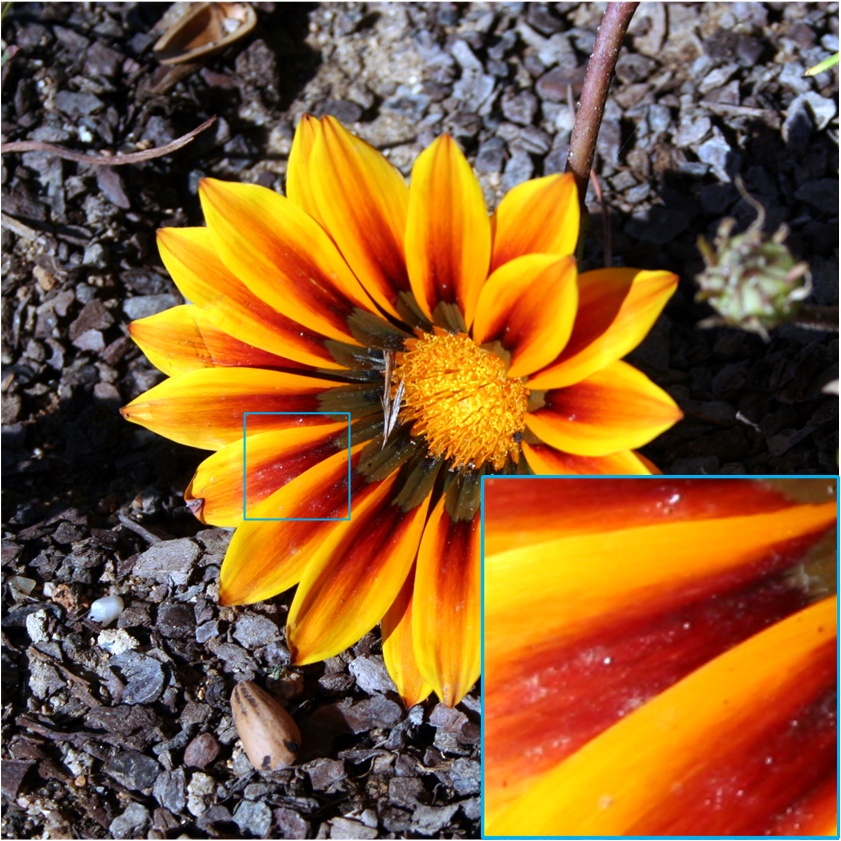}
\end{minipage}
}%
\subfigure[JPEG(0.14/23.05/0.85)]{
\begin{minipage}[t]{0.33\linewidth}
\centering
\includegraphics[scale=0.25]{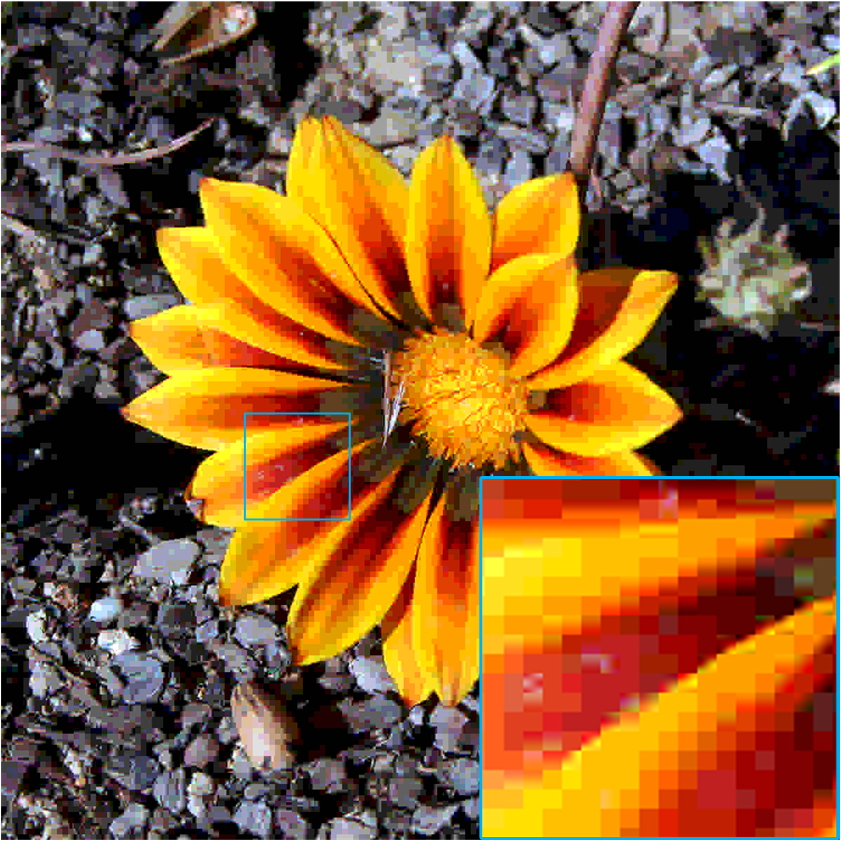}
\end{minipage}
}%
\subfigure[JPEG2000(0.14/28.66/0.955)]{
\begin{minipage}[t]{0.33\linewidth}
\centering
\includegraphics[scale=0.25]{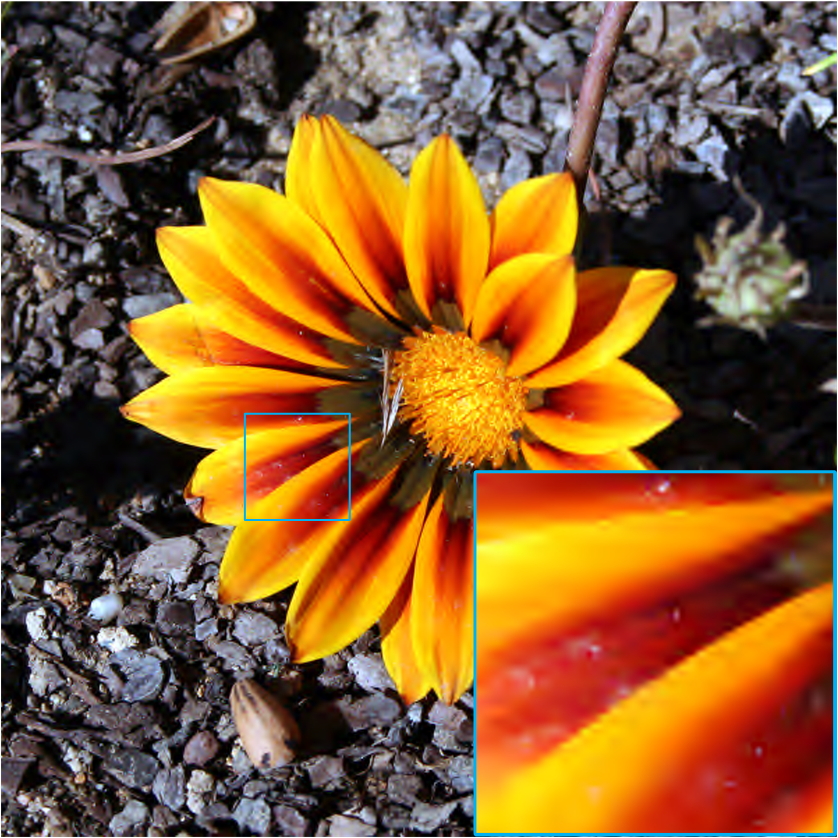}
\end{minipage}
}%

\subfigure[BPG(0.122/29.09/0.956)]{
\begin{minipage}[t]{0.33\linewidth}
\centering
\includegraphics[scale=0.25]{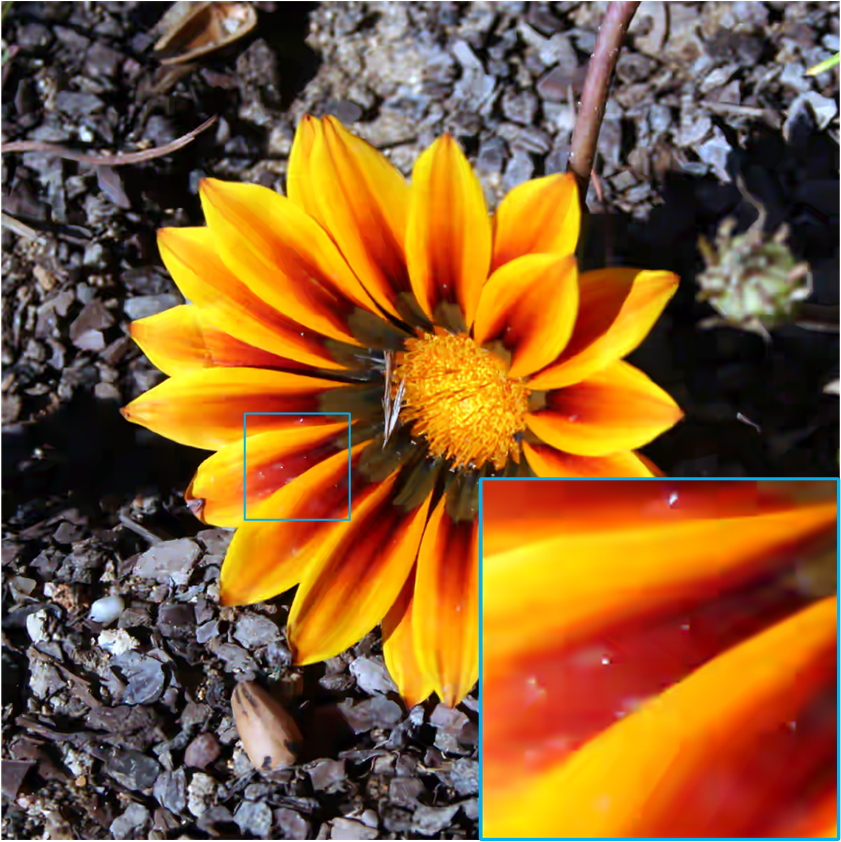}
\end{minipage}
}%
\subfigure[DSSLIC(0.130/29.45/0.957)]{
\begin{minipage}[t]{0.33\linewidth}
\centering
\includegraphics[scale=0.25]{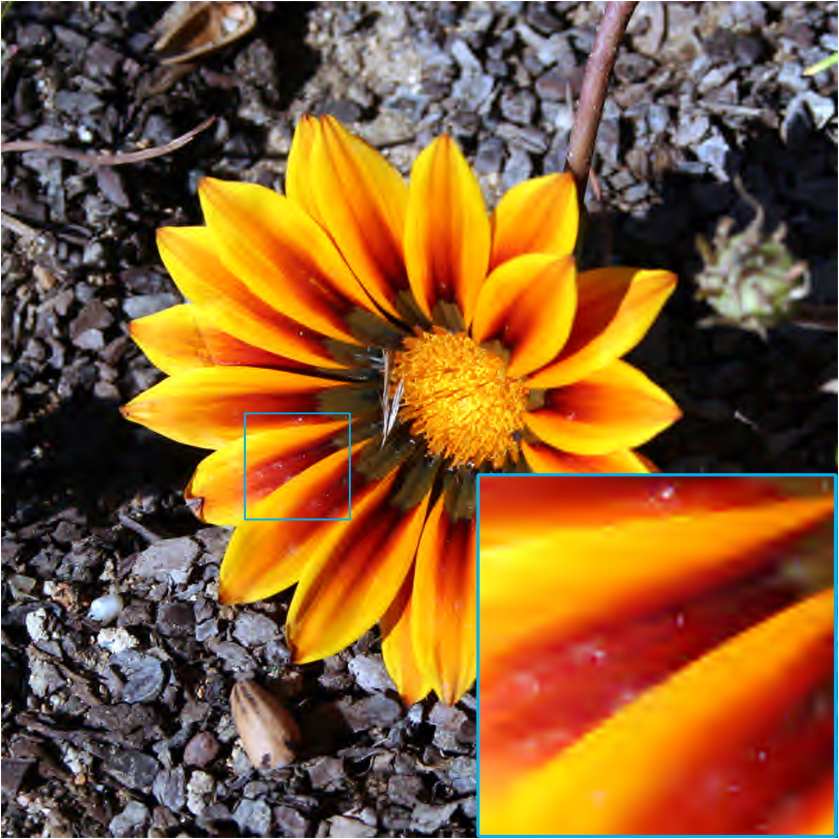}
\end{minipage}
}%
\subfigure[Ours(0.128/30.6/0.960)]{
\begin{minipage}[t]{0.33\linewidth}
\centering
\includegraphics[scale=0.25]{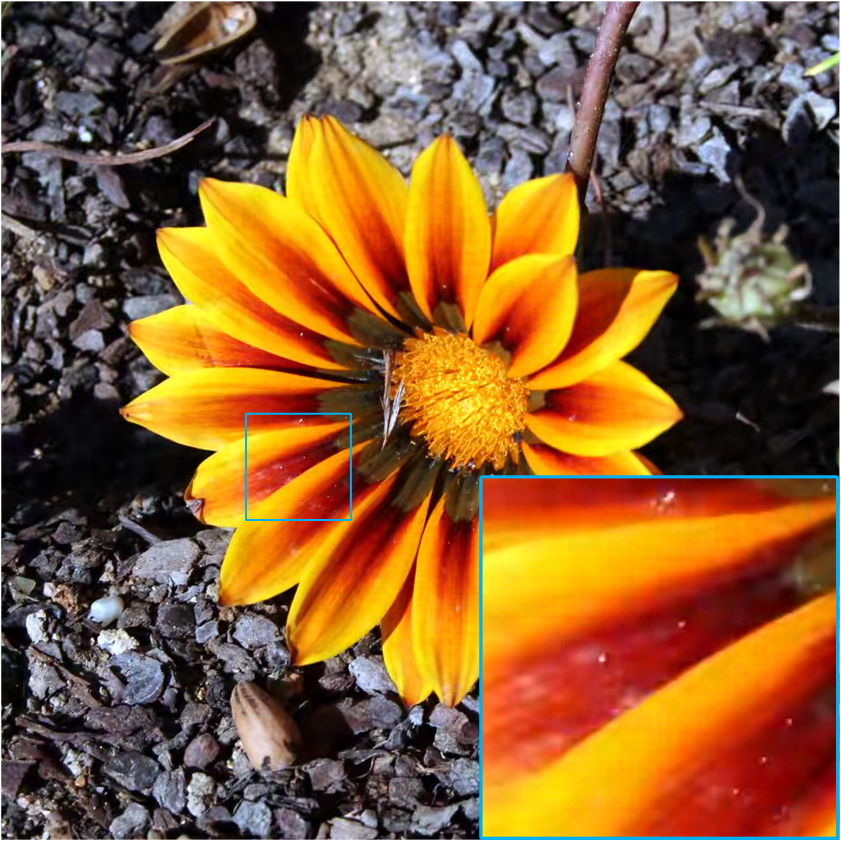}
\end{minipage}
}%
\centering
\caption{Example 2 in the tecnick dataset (bits/pixel/channel, PSNR, MS-SSIM).}
\label{Example_2}
\end{figure}

To compare the visual quality of different methods, we show some examples from the Kodak dataset in Fig. \ref{Example_1} to Fig. \ref{Example_2}, with some regions amplified. In the test, the shift scaling method is used in our method. By analyzing all the samples, we can get the following conclusions. Compared to other codecs, the JPEG has the worst performance due to the blocking artifacts. The JPEG2000 shows some ringing artifacts. The WebP codec has better performance than JPEG2000, but it is blurry in certain areas of the image. The BPG result is quite smooth, but it loses some detailed information. DSSLIC is significantly better than BPG, and our method achieves even better result than DSSLIC.

\begin{figure}
\centering
\subfigure{
\begin{minipage}[t]{0.5\linewidth}
\centering
\includegraphics[width=\columnwidth]{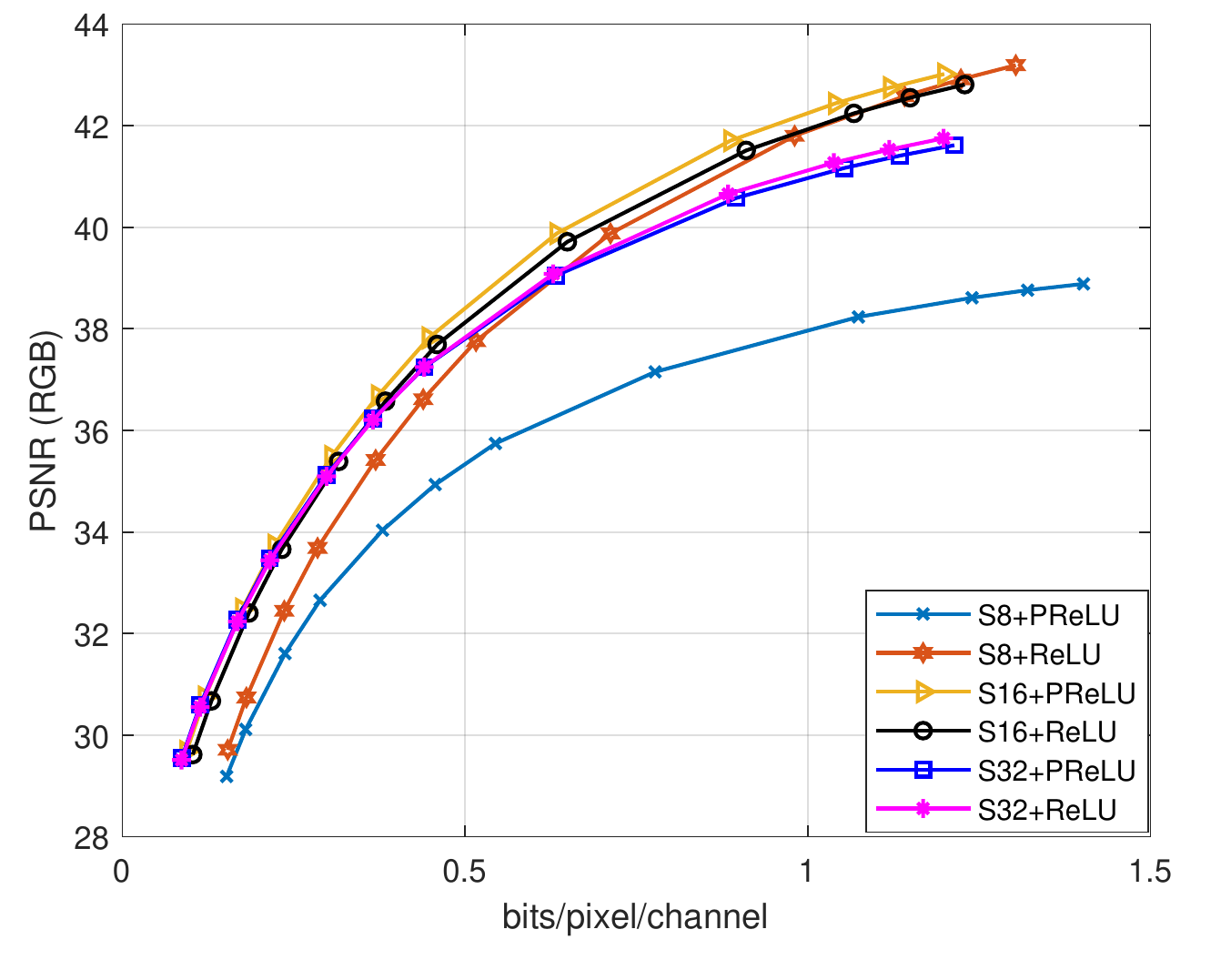}
\end{minipage}
}%
\subfigure{
\begin{minipage}[t]{0.5\linewidth}
\centering
\includegraphics[width=\columnwidth]{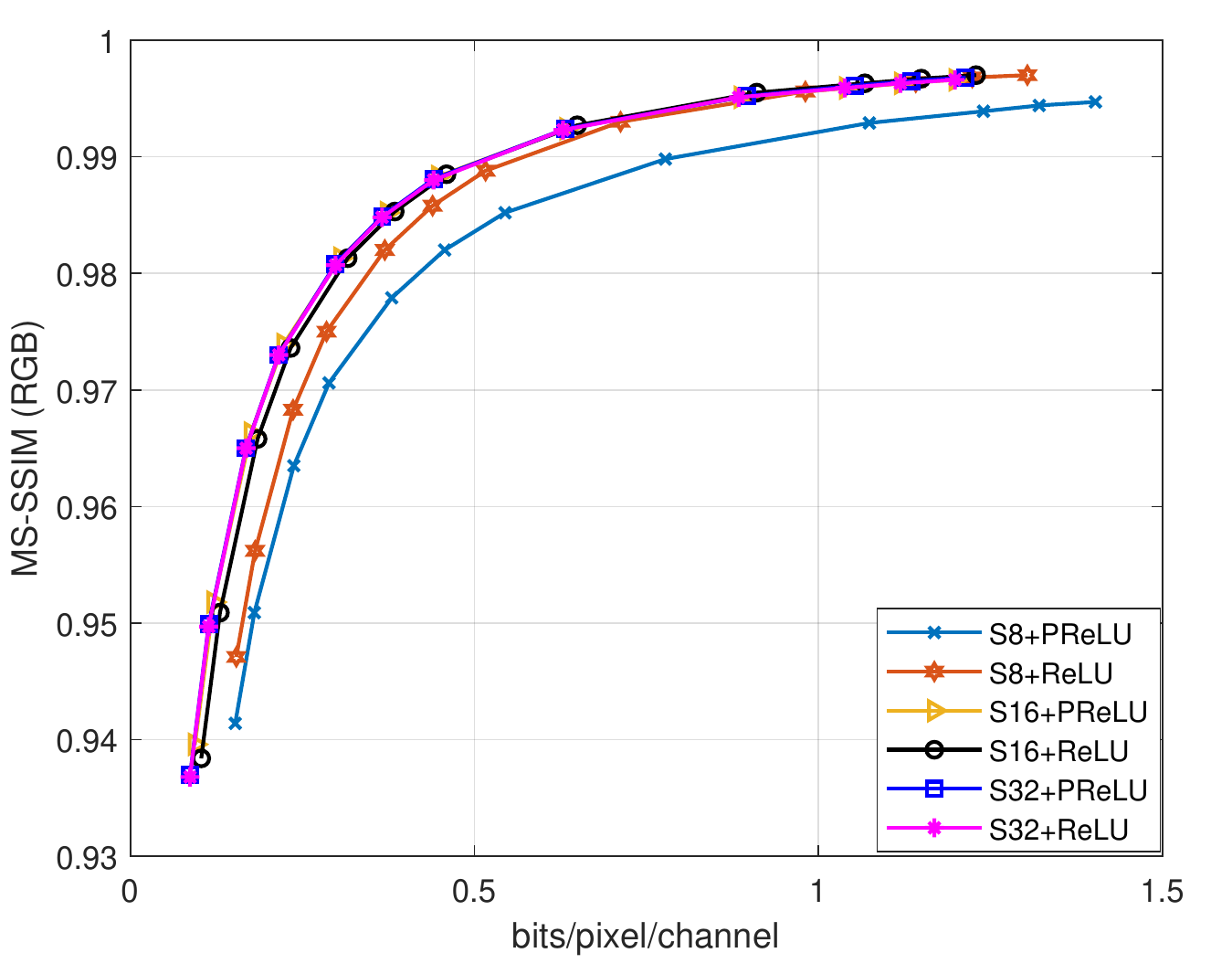}
\end{minipage}
}%
\centering
\caption{The effect of different layers and activation functions on the performance of the framework on all 24 Kodak images in terms of PSNR and MS-SSIM.  \textbf{Sn:} the width and the height of the compact image $c$ are both $1/n$ of those of the input image. \textbf{PReLU or ReLU:} the activation functions of the network.}
\label{Compare_result}
\end{figure}

\subsection{Other Ablation Studies}

We also conduct ablation studies to investigate the impacts of the size of the compact representation and the activation functions in our framework. The result is shown in Fig. \ref{Compare_result}. It can be seen that the best results are given when the width and height of the compact image $c$ are both $1/16$ of those of the original image, compared to $1/8$ and $1/32$. In this case, PReLU yields slightly better performance than ReLU. Therefore the configuration of S16+PreLU is selected in our experimental results above. In addition, the PSNR is more susceptible to the compact map size and activation function, compared to the SSIM.

\section{Conclusion}

In this paper, we propose an improved hybrid image compression framework, which is a simplified version of the state-of-the-art DSSLIC scheme in \cite{8683541} without using the semantic segmentation layer. Various modifications to the autoencoder network are proposed, and a new clipping method is proposed to scale the residual image. The performance of our method in both PSNR and MS-SSIM metrics is better than BPG codec and DSSILC across a large range of bit rates. In the future work, we will further improve the performance of the image on the MS-SSIM metric, and we will investigate how to encode the residual image more efficiently.

\section*{Acknowledgment}

This work was supported by the National Natural Science Foundation of China under Grant No. 61474093 and by Tencent.

\bibliographystyle{unsrt}  


\end{document}